\documentclass[pre,aps,10pt,amsmath,amsfonts,amssymb,twocolumn]{revtex4-2}
\usepackage{graphicx,CJK,bm}
\newcommand{\aslim}{\stackrel{\text{a.s.}}{\longrightarrow}}
\DeclareMathOperator{\Tr}{Tr}
\begin{document}
\title{Physical meaning of principal component analysis for classical lattice systems with
translational invariance}
\begin{CJK*}{UTF8}{mj}
\author{Su-Chan Park (박수찬)}
\email{spark0@catholic.ac.kr}
\affiliation{Department of Physics, The Catholic University of Korea, Bucheon 14662, Republic of Korea}
\date{\today}
\begin{abstract}
We explore the physical implications of applying principal component analysis (PCA) to translationally invariant classical systems defined on a $d$-dimensional hypercubic lattice. Using Rayleigh-Schr\"odinger perturbation theory, we demonstrate that the principal components are related to the reciprocal lattice vectors of the hypercubic lattice, and the corresponding eigenvalues are connected to the discrete Fourier transform of the sampled configurations. From a different perspective, we show that the PCA in question can be viewed as a numerical method for computing the ensemble average of the squared moduli of the Fourier transform of physical quantities. Our results also provide a way to determine approximately the principal components of a classical system with translational invariance without the need for matrix diagonalization.
\end{abstract}
\maketitle
\end{CJK*}
\section{introduction}
An unprecedented achievement of machine learning has brought physics into
the realm of data-driven science~\cite{Mehta2019,Carleo2019}. 
In statistical and condensed-matter physics, machine learning has been successfully applied
to classify classical~\cite{Wang2016,Wetzel2017,Car2017,Hu2017,Wang2017,WZ2018,Kiwata2019,Sale2022}, 
quantum~\cite{ZK2017,Chng2018,Yang2021}, 
and nonequilibrium~\cite{MCL2019,SLDZ2021,Shen2022,Tang2024,muzzi2024} phases of matter.  
Despite the numerous successful applications of machine learning to physics problems, 
a comprehensive theoretical understanding of this success remains lacking. 
In this sense, machine learning presents a theoretical challenge.

Given the complexity of successful machine learning algorithms, establishing a theoretical framework 
for these methods seems a formidable task. 
As such, a promising approach would be to first gain a theoretical understanding of simpler cases.
One such candidate is principal component analysis 
(PCA)~\cite{Jolli2002, Jolliffe2016}, 
a widely used unsupervised learning algorithm that focuses on the singular values of 
a matrix constructed from data. 
Although the full theoretical treatment of PCA is still challenging, 
understanding its application to more specific problems can be a more tractable goal. 

In this context, we explore the application of PCA to certain systems with a 
specific, yet quite general, property within physics. Specifically, we examine PCA applied to 
translationally invariant classical systems undergoing phase transitions, 
where it has shown potential for identifying transition points across a range of 
systems~\cite{Wang2016, Wetzel2017, Car2017, Hu2017, Wang2017, WZ2018, Yang2021, Sale2022, muzzi2024}. 
This is the central focus of the present paper.
Since PCA applied to translationally invariant classical systems is the primary focus of this paper, 
we refer to this approach simply as ``the PCA'' henceforth.

The Fourier transform plays a crucial role in translationally invariant systems. 
In this regard, it may be natural to hypothesize that singular values obtained 
through the PCA should be related with the Fourier transform of the system in one way or another. 
Indeed, some numerical observations in simple systems, such as the Ising model, have been found to be 
consistent with this hypothesis~\cite{Wang2016, Hu2017, WZ2018}.
To our knowledge, however, no systematic theoretical treatment of this relation exists in the literature.
In this paper, we theoretically show that the 
PCA and the Fourier transform are indeed intimately related. 
This relation enables us to assess whether the PCA can be effective in studying phase transitions 
by examining the relationship between the Fourier transform and the order parameter, 
once the latter is already known. Conversely, if the PCA successfully classifies the phases of matter 
in a system where the order parameter is unknown, then one can identify the order parameter 
as one of the Fourier modes.

This paper is organized as follows. 
In Sec.~\ref{Sec:met}, we establish the connection between
Fourier modes and the singular values obtained through PCA, specifically applied to translationally
invariant classical systems on a $d$-dimensional hypercubic lattice.
Although we limit ourselves to the $d$-dimensional hypercubic lattice, 
our theory can be easily extended to other Bravais lattices.
This section is divided into two subsections. In Sec.~\ref{Sec:mss}, 
we establish the mathematical framework for data that are fed to the PCA algorithm.
Here, we outline the general form of the data and the key assumptions. 
We also articulate what is meant by the translational invariance in this paper.
In Sec.~\ref{Sec:an}, 
we find the connection between the principal components of the data and the Fourier 
transform. Specifically, we demonstrate that discrete Fourier tranforms of configurations 
provide good approximations for all singular values of the data.
In Sec.~\ref{Sec:num}, 
we analyze example systems, 
to support the theory developed in Sec.~\ref{Sec:met}. 
These examples not only confirm the predictions of the theory 
but also help us interpret the singular values obtained from the PCA in a more physical context.
In Sec.~\ref{Sec:sum}, we summarize the main findings of the paper.

\section{\label{Sec:met}method}
A typical PCA study on phase transitions involves two main steps.
In the first step, referred to as the data-preparation step in this paper 
(and detailed in Sec.~\ref{Sec:mss}), 
independent configurations are sampled using Monte Carlo simulations.
The data of configurations are organized into a matrix by concatenating the configurations.
The second step, referred to as the PCA step in this paper
(and detailed in Sec.~\ref{Sec:an}), 
finds singular values of the matrix prepared in the data-preparation step.
It is the largest singular value (corresponding to the leading principal component) 
that identifies the critical point and helps estimate a certain critical exponent~\cite{Wang2016,Car2017,Hu2017,Sale2022,muzzi2024}.

\subsection{\label{Sec:mss}Mathematical framework: The data-preparation step}
We assume that a system under consideration is defined on a $d$-dimensional hypercubic lattice of size 
$L = \prod_{\ell=1}^d L_\ell$ with periodic boundary conditions. 
Each site is designated as a $d$-dimensional lattice vector $\bm{a}=(n_1,\ldots,n_d)$ 
with $1 \le n_\ell \le L_{\ell}$ ($\ell=1, 2,\ldots,d$).
For a later reference, we define a set ${\cal L}$ of all lattice vectors as
$$
{\cal L} := \{(n_1,\ldots,n_d): 1 \le n_\ell \le L_{\ell}\}.
$$
Throughout the paper, we denote a vector by a boldface italic letter like $\bm{a}$, $\bm{b}$, $\bm{c}$,
$\bm{p}$, $\bm{q}$.

At every site $\bm{a}$, a random variable $\xi_{\bm{a}}$ is assigned. 
The physical meaning of $\xi_{\bm{a}}$ depends on what system is under consideration.
For the Ising model, $\xi_{\bm{a}}$ is a spin variable that takes either $-1$ or $1$.
For spin-$1$ models like the Blume-Capel model~\cite{Bl,Capel}, 
$\xi_{\bm{a}}$ is a spin variable that takes $-1$, $1$, or $0$.
For a hard-core classical particle system, $\xi_{\bm{a}}$ 
takes either $1$ or $0$ to represent that a site is occupied or empty. 
For a classical boson system, $\xi_{\bm{a}}$ takes a nonnegative integer that counts the number of
particles at site ${\bm{a}}$ (see Ref.~\cite{P2005,P2006} for a numerical method to simulate 
classical reaction-diffusion systems of bosons). 
For the $n$-vector model~\cite{Stanley1968} with $n\ge 2$, $\xi_{\bm{a}}$ 
is a continuous $n$-dimensional classical spin with modulus one. 

By a configuration (of a system under consideration), we mean a (ordered) collection of 
all $\xi_{\bm{a}}$'s. We will denote a configuration by $C$.
For the PCA to work, we need probability $P(C)$ of being in configuration $C$.
In the case that $\xi_{\bm{a}}$ takes continuous values, 
$P(C)$ should be understood as a probability density. 
According to how a problem is defined, $P(C)$ is given either by an explicit
formula or by an implicit solution of a time-evolution equation.
If one is interested in an equilibrium system with inverse temperature $\beta$, 
then $P(C)$ is explicitly given as
\begin{align}
\label{Eq:equil}
P(C) = \frac{1}{Z} e^{-\beta H(C)},
\end{align}
where $H(C)$ is the Hamiltonian in question and $Z$ is the partition function.
If a system is defined by the master equation,
then $P(C)$ is implicitly given by the solution of the master equation at certain fixed time 
(including $\infty$, if a stationary state is well defined).
In Sec.~\ref{Sec:num}, we will study examples of the above two cases.
For a later reference, we define the ``ensemble'' average of a random variable, say, $Y$ as
$$
\langle Y \rangle := \sum_C Y(C) P(C),
$$
where $Y(C)$ means a realization of $Y$ in configuration $C$ and $\sum$ should be understood as an 
integral if $P(C)$ is a probability density.

When we write $C_\alpha$ for a configuration, the random variable at site $\bm{a}$ in $C_\alpha$
will be denoted by $\xi_{\bm{a}}^{(\alpha)}$. When we say that a system has 
the translational invariance, we mean $P(C_1) = P(C_2)$
if there exists a vector $\bm{c}$ for two configurations $C_1$ and $C_2$ such that 
$\xi_{\bm{a}}^{(1)}=\xi_{\bm{a} + \bm{c}}^{(2)}$ for every lattice vector $\bm{a}
\in {\cal L}$ 
up to periodic boundary conditions.

In the data-preparation step, 
$N$ configurations are independently drawn from $P(C)$ using Monte Carlo simulations.
Needless to say, the explicit mathematical formula for $P(C)$ is not indispensable,
as long as the configurations can be collected. To be formal, 
let $C_{\alpha}$'s be independent and identically distributed $N$ random 
configurations sampled from the common probability distribution $P(C)$ ($\alpha = 1,2,\ldots,N$). 
Note that samples of $C_{\alpha}$'s are obtained in the data-preparation step.
For a later reference, we define the ``empirical'' average of a random variable Y as
\begin{align}
\label{Eq:empave}
[ Y ]_N := \frac{1}N \sum_{\alpha=1}^N Y({C_\alpha}).
\end{align}
Obviously, the empirical average of $Y$ is a random variable different from $Y$.
From time to time, we will also regard $[Y]_N$ as a sequence of random variables.
If the variance of $Y$ is finite just as most of quantities in finite physical systems,
the strong law of large numbers~\cite{FellerI} guarantees that 
$[Y]_N$ converges almost surely to $\langle Y \rangle$ under the infinite $N$ limit.
Representing the almost sure limit of a sequence of random variables 
by an arrow above which a text ``a.s.'' is added, we write
$$
[Y]_N \aslim \langle Y \rangle.
$$
For the definition of the almost sure limit, see, for example, Chapter 2 of Ref.~\cite{Gardiner}. 
The strong law of large numbers will play an important role in developing the theory in the next subsection.

Data one feeds to the PCA algorithm 
are normally prepared in a form of $N \times L$ matrix $X$ in which the $\alpha$th row 
contains information of the sampled $C_\alpha$ ($\alpha=1,2,\ldots,N$).
Since a $d$-dimensional configuration should be put into a row of $X$, 
we need a one-to-one mapping $v : {\cal L} \mapsto \{1,2,\ldots, L\}$
from the set ${\cal L}$ of all lattice vectors to the set $\{1,2,\ldots,L\}$ of $L$ integers.
In what follows, we will refer to $v$ as a vector-to-integer (VI) mapping.
One example of a VI mapping is
\begin{align}
\label{Eq:defn}
v(\bm{a}) = n_1 + \sum_{\ell=2}^{d} \left ( \prod_{i=1}^{\ell-1} L_i \right ) (n_\ell-1),
\end{align}
which is $v(\bm{a}) = n_1$ in one dimension, $v(\bm{a}) = n_1 + L_1(n_2-1)$ in two dimensions, 
$v(\bm{a})=n_1+L_1(n_2-1)+L_1L_2(n_3-1)$ in three dimensions, and so on.
In Sec.~\ref{Sec:num}, we will use Eq.~\eqref{Eq:defn} for numerical calculations.

As we will see soon, however, the exact form of a VI mapping is immaterial in
developing our theory. Actually, 
the purpose of introducing a VI mapping just lies in completing the mathematical framework.
For convenience, we assume that a VI mapping is fixed in the following.

For the given VI mapping, the $(\alpha,m)$ entry of $X$ is given as
$$X_{\alpha m}= x_m^{(\alpha)} := f\left (\xi_{\bm{a}}^{(\alpha)}\right ),$$ 
where $\bm{a}$ is the preimage of $m$ of the VI mapping, that is, $m=v(\bm{a})$, and 
$f$ is a certain function, which may depend on the lattice vector. 
Since $f$ is normally not introduced in the literature, we feel it necessary to explain
the role of $f$ by examples.
The identity function $f(\xi)=\xi$ would be the most natural choice, but
other choices can also be meaningful.
For the Ising model ($\xi_{\bm{a}} = \pm 1$),
the identity function is (implicitly) used in Ref.~\cite{Wang2016}.
Still, one may choose $f(\xi) = (1+\xi)/2$ to study the Ising model 
as a lattice-gas model.
This lattice-gas version will be discussed in Sec.~\ref{Sec:ising}.
An example of a lattice-vector dependent $f$ will be given in Sec.~\ref{Sec:ising} in the context 
of the antiferromagnetic Ising model. 
The choice of $f(\xi) = \xi^2$ is useless in the Ising model,
but this function can be helpful in the study of a spin-1 model~\cite{Hu2017}.
For the $XY$ model in which $\xi_{\bm{a}}$ is a planar vector with modulus one,
one can choose $f(\xi) = \xi_{(1)}+i\xi_{(2)}$,
where $\xi_{(j)}$ is the $j$th component of the spin vector.
This example shows that $x_m^{(\alpha)}$ can be a complex number.
For a bosonic reaction-diffusion system with $\xi_{\bm{a}}$ to take nonnegative integer,
one may choose $f(\xi) = 1 - \delta_{\xi,0}$, where
$\delta$ is the Kronecker $\delta$ symbol.
This choice can be useful if one is only interested in whether a site is
occupied or not. As the above examples show, a physical intuition is in a sense implicitly 
added to a PCA study through the (appropriate) choice of $f$; more detailed discussion can be 
found in Sec.~\ref{Sec:ising}.

Now, $X$ that will be a playground of the PCA has the form
\begin{align*}
X = 
\begin{pmatrix}
x_1^{(1)}&x_2^{(1)}&\cdots&x_L^{(1)}\\
x_1^{(2)}&x_2^{(2)}&\cdots&x_L^{(2)}\\
\vdots&\vdots&\vdots&\vdots\\
x_1^{(N)}&x_2^{(N)}&\cdots&x_L^{(N)}
\end{pmatrix}.
\end{align*}
A centered matrix $\overline{X}$ with components
$$
\overline{X}_{\alpha m} = \overline{x}_m^{(\alpha)}:=x_m^{(\alpha)}- [x_m]_N
$$ 
can also be fed to the PCA algorithm.
Recall the definition~\eqref{Eq:empave} of the empirical average $[x_m]_N$.
The PCA amounts to finding the singular values of $X$ and $\overline{X}$.

\subsection{\label{Sec:an}Perturbation theory for the singular values: The PCA step}
In the PCA step, the singular values of $X$ and $\overline{X}$ are calculated.
More precisely, the main interest of the PCA is the spectral decomposition of 
an empirical second-moment matrix $M$ and an empirical covariance matrix $G$, defined as
\begin{align*}
M := \frac1N X^\dag X, \quad G := \frac1N \overline{X}^\dag \overline{X}. 
\end{align*}
The elements of $M$ and $G$ can be written as 
\begin{align*}
M_{nm} &= \frac{1}{N} 
\sum_{\alpha=1}^{N} x^{(\alpha)*}_n x^{(\alpha)}_m
= \left [ x_n^* x_m \right ]_N,\\
G_{nm} &= \frac{1}{N} 
\sum_{\alpha=1}^{N} \overline{x}^{(\alpha)*}_n \overline{x}^{(\alpha)}_m\\
&=
\left [ \left ( x_n  - [x_n]_N\right )^*
\left ( x_m  - [x_m]_N\right )\right ]_N,
\end{align*}
where the asterisk means complex conjugate.
If $x$'s are vectors as in the $n$-vector model with $f(\xi)=\xi$, 
then the multiplications $x_n^* x_m$ should be understood as an inner product of two vectors.
Obviously, $M$ and $G$ are positive semidefinite matrices.
We will denote the $r$th largest eigenvalue 
of $M$ and $G$ by $\lambda_r^M$ and $\lambda_r^G$, respectively, and will refer to the $r$ 
as ranking.

Let us find formulas that approximate all $\lambda$'s. 
For presentation, we found it most convenient to write $x_{\bm{a}}$ for 
$x_m$ with $m=v(\bm{a})$.
Similarly, we also write a component of an $L\times L$ matrix, for example, $M$ as $M_{\bm{a}\bm{b}}$,
meaning
$$
M_{\bm{a}\bm{b}} \equiv M_{v(\bm{a})v(\bm{b})}.
$$ 
We introduce positive semidefinite matrices $S$ and $\overline{S}$ with elements
\begin{align*}
S_{\bm{a}\bm{b}} &= \frac{1}{L} \sum_{\bm{c} \in {\cal L}} 
\left [ x_{\bm{a}+\bm{c}}^*\, x_{\bm{b}+\bm{c}} \right ]_N,\\
\overline{S}_{\bm{a}\bm{b}} &= \frac{1}{L} \sum_{\bm{c} \in {\cal L}} 
\left [ \overline{x}_{\bm{a}+\bm{c}}^*\, \overline{x}_{\bm{b}+\bm{c}} \right ]_N,
\end{align*}
where periodic boundary conditions are assumed. 
Both $S$ and $\overline{S}$ are translationally invariant in that 
\begin{align}
\label{Eq:Str}
S_{\bm{a} + \bm{e}_\ell,\bm{b} + \bm{e}_\ell} =S_{\bm{a}\bm{b}},\quad
\overline{S}_{\bm{a} + \bm{e}_\ell,\bm{b} + \bm{e}_\ell} =\overline{S}_{\bm{a}\bm{b}},
\end{align}
where ${\bm e}_\ell$ is the unit vector along the $\ell$th direction ($\ell=1,2,\ldots,d$).
Obviously,
\begin{align}
\label{Eq:tr}
\Tr S &= \Tr M = \sum_{\bm{a}\in{\cal L}} \left [|x_{\bm{a}}|^2 \right ]_N=\sum_{r=1}^L \lambda_r^M,\\
\Tr \overline{S} &= \Tr G = \sum_{\bm{a}\in{\cal L}} \left [|\overline{x}_{\bm{a}}|^2 \right ]_N=\sum_{r=1}^L \lambda_r^G.
\nonumber
\end{align}

We define an $L\times 1$ matrix $|\bm{p}\rangle$ for each reciprocal lattice vector
($0 \le k_\ell \le L_\ell - 1$)
$$
\bm{p} = 2\pi \left ( \frac{k_1}{L_1},\frac{k_2}{L_2},\ldots,\frac{k_d}{L_d}\right ),
$$
as
$$
|\bm{p} \rangle := \frac{1}{\sqrt{L}} \sum_{\bm{a}\in {\cal L}} \exp\left ( i \bm{p} \cdot \bm{a}
\right ) |\bm{a}\rangle_v,
$$
where $|\bm{a}\rangle_v$ is an $L\times 1$ matrix with one at the $v(\bm{a})$th row and zeros at
all the other rows.
Obviously, the set of all $|\bm{p}\rangle$'s is an orthonormal basis of the $L$-dimensional
inner product space spanned by all $|\bm{a}\rangle_v$'s.

By Eq.~\eqref{Eq:Str}, every $|\bm{p}\rangle$ is
the eigenstate of both $S$ and $\overline{S}$ 
with the corresponding eigenvalues
\begin{subequations}
\label{Eq:theory}
\begin{eqnarray}
\label{Eq:theorya}
\tau_{\bm{p}}^M&:=&
\langle \bm{p} | S | \bm{p} \rangle = 
\left [ \left | \tilde x_{\bm{p}} \right |^2 \right ]_N,\\
\tau_{\bm{p}}^G&:=&
\langle \bm{p} | \overline{S} | \bm{p} \rangle = 
\left [ \left | \tilde x_{\bm{p}} -\left [\tilde x_{\bm{p}}\right ]_N\right |^2 
\right ]_N,
\label{Eq:theoryb}
\end{eqnarray}
\end{subequations}
where $\langle \bm{p} | := | \bm{p} \rangle^\dag$ as usual and
\begin{align*}
\tilde x_{\bm{p}} &:=  \frac{1}{\sqrt{L}}\sum_{\bm{a}}e^{i \bm{p} \cdot \bm{a}} 
x_{\bm{a}} 
\end{align*}
is the discrete Fourier transform of $x_{\bm{a}}$. 
Note that Eq.~\eqref{Eq:theory} does not depend on the choice of a VI mapping.
Let
\begin{align}
\label{Eq:exact}
\varphi_{\bm{p}}^M:=\langle |\tilde{x}_{\bm{p}}|^2\rangle,\quad
\varphi_{\bm{p}}^G:=\langle |\tilde{x}_{\bm{p}} - \langle \tilde{x}_{\bm{p}} \rangle  |^2\rangle.
\end{align}
Since $\tau_{\bm{p}}^{M,G} \aslim \varphi_{\bm{p}}^{M,G}$,
we can regard $\tau_{\bm{p}}^{M,G}$ as a numerical approximation 
of $\varphi_{\bm{p}}^{M,G}$ by Monte Carlo simulations.

Let $\delta M := M - S$ and $\delta G:= G- \overline{S}$. 
Since
$
\langle S_{nm}\rangle = \langle M_{nm} \rangle$ and 
$\langle \overline{S}_{nm}\rangle = \langle G_{nm}\rangle$
due to the translational invariance, 
we have 
$
\langle \delta M_{nm} \rangle = 
\langle \delta G_{nm} \rangle = 
 0
$ for all $n$, $m$.
By the strong law of large numbers, therefore, both $\delta M$ and $\delta G$ should 
converge almost surely to the zero matrix under the infinite $N$ limit.
If $N$ is sufficiently large, then 
$\delta M$ and $\delta G$ can be treated as a small perturbation and
we can resort to the Rayleigh-Schr\"odinger perturbation theory in quantum mechanics
to approximately find all the eigenvalues of $M$ and $G$.

Up to the first order, we have the approximate eigenvalues of $M$ and $G$ as 
$\langle \bm{p} |M |\bm{p}\rangle$ and
$\langle \bm{p} |G |\bm{p}\rangle$, respectively,
which are actually identical to $\tau_{\bm{p}}^{M}$ and
$\tau_{\bm{p}}^{G}$ in Eq.~\eqref{Eq:theory}.
Denoting the $r$th largest value among all 
$\tau_{\bm{p}}^{M,G}$'s ($\varphi_{\bm{p}}^{M,G}$'s) by $\tau_r^{M,G}$
($\varphi_r^{M,G}$), we finally have
\begin{align}
\label{Eq:theory2}
\varphi_r^M \approx\lambda_r^M \approx \tau_r^M,\quad
\varphi_r^G \approx\lambda_r^G \approx \tau_r^G.
\end{align}
Therefore, one can find all the eigenvalues of $M$ and $G$ approximately using Eq.~\eqref{Eq:theory}.
In other words,
we found that the Fourier transform $\tilde x_{\bm{p}}$ is  intimately 
related with the principal components in question. Furthermore, we can say that the approximation
becomes more and more accurate as $N$ increases in that $\tau_r^{M,G}-\lambda_r^{M,G} \aslim 0$.

Equation~\eqref{Eq:theory2} gives an insight that
the PCA can be regarded as a numerical method to calculate $\varphi_{\bm{p}}^{M,G}$ approximately.
In this context, whether or not the PCA can be a good tool to study a phase transition depends on
the physical meaning of $\varphi_1^{M,G}$.
If either $\varphi_1^M$ or $\varphi_1^G$ (or both) happens to be related to the order parameter
of the phase transition in question, then the PCA has potential to find the critical point. 
Otherwise, the PCA would fail.

\section{\label{Sec:num}Applications}
In this section, we numerically support Eq.~\eqref{Eq:theory2} using 
two concrete examples.  One is the one-dimensional contact process (CP)~\cite{H1974} and the other
is the two-dimensional Ising model.
In due course, we associate $\lambda_1^{M}$ and $\lambda_1^{G}$ with physical quantities 
that are usually studied in the context of phase transitions.
After these numerical studies, discussion about the $q$-states Potts model and the $n$-vector model 
is followed.

Since the purpose of this paper is to find a relation between 
the largest singular value of, for example, $X$ and the order parameter
of a system under consideration, 
we do not investigate critical phenomena in detail.
Needless to say, we always assume periodic boundary conditions.

\subsection{\label{Sec:cp}One dimensional contact process}
In this subsection, we apply PCA to the CP, which is regarded as an Ising model of 
absorbing phase transitions. 
One may consult, e.g., Refs.~\cite{H2000,O2004} for a review of absorbing phase transitions. 
After numerically confirming Eq.~\eqref{Eq:theory2} for a specific case of the CP, we will 
discuss the physical meaning of the PCA, which also provides an understanding about the success 
of a recent PCA study~\cite{muzzi2024} on systems that exhibit absorbing phase transitions. 

For completeness, we briefly explain the dynamic rule of the CP in $d$ dimensions.
We use notations in a consistent way with Sec.~\ref{Sec:met}.
$\xi_{\bm{a}}$ takes either zero or one. If $\xi_{\bm{a}}=1$, then we say that a particle
is located at site ${\bm{a}}$; no multiple occupancy is allowed. 
If $\xi_{\bm{a}}=0$, then we say that site ${\bm{a}}$ is empty.
With transition rate $p$, a particle is removed from the system. 
With transition rate $1-p$, a particle attempts to branch an offspring 
to one of its nearest-neighbor sites that is chosen at random.
We only consider the fully occupied initial condition ($\xi_{\bm{a}}=1$ for all ${\bm{a}} \in {\cal L}$ at $t=0$)
for Monte Carlo simulations, which ensures the translational invariance
for all time $t$. 

The CP is known to have the critical point $p_c$ in all dimensions.
When $p$ is larger than $p_c$, 
particles tend to die out and the system falls into
the absorbing state (the configuration without a particle) in short time. 
It is said that the system is in the absorbing phase when $p>p_c$. 
When $p$ is smaller than $p_c$, density of occupied sites can remain finite for all time $t$
(strictly speaking, the steady state with nonzero density exists only when infinite $L$ limit is taken
before the infinite time limit). 
It is said that the system is in the active phase when $p<p_c$. 
Since the density is zero in the absorbing phase and nonzero in the active phase,
the density is the order parameter of the absorbing phase transition.

Since we are only interested in the validity of Eq.~\eqref{Eq:theory2},
we simulated a one-dimensional system with small size ($L=64$) up to short time ($t=40$). 
We also set $p=0.232$.
In this case, $P(C)$ is the solution of the master equation at $t=40$.
In the data-preparation step, we performed Monte Carlo simulations using 
the following algorithm. 
Assume that there are $K$ particles in the system at time $t$.
One particle is chosen at random. With probability $p$, it is removed
from the system. With probability $1-p$, it branches an offspring to one of its nearest neighbors 
that is chosen at random. If the offspring happens to be placed at a site already occupied, 
then the offspring is immediately removed and effectively no configuration change occurs. 
After the above probabilistic attempt, time increases by $1/K$.
The simulation is terminated if no particle is left or time exceeds the preassigned
observation time (40, in this example). In the data we will present soon, at least one particle is left
in all sampled configurations at $t=40$.

In the data-preparation step, we performed $10^4$ independent simulations up to $t=40$.
By setting $x_n = \xi_n$, we have $X$ in the form 
$$
X = \begin{pmatrix} 
1&0&0&\cdots&1\\
1&0&1&\cdots&0\\
0&0&1&\cdots&0\\
\vdots&\vdots&\vdots&\vdots&\vdots\\
0&1&0&\cdots&1\\
\end{pmatrix}.
$$
To investigate the effect of $N$, we constructed three different $X$'s with 
$N=10^2$, $10^3$, and $10^4$ rows.
In the PCA step, we performed numerical matrix diagonalization of $M$ and $G$, 
using the method \verb|eigvalsh|
in the \verb|numpy.linalg| package of {\bf Python}. 
For convenience, the eigenvalues obtained by the numerical matrix diagonalization will be
said to be ``exact.''

\begin{figure}
\includegraphics[width=\linewidth]{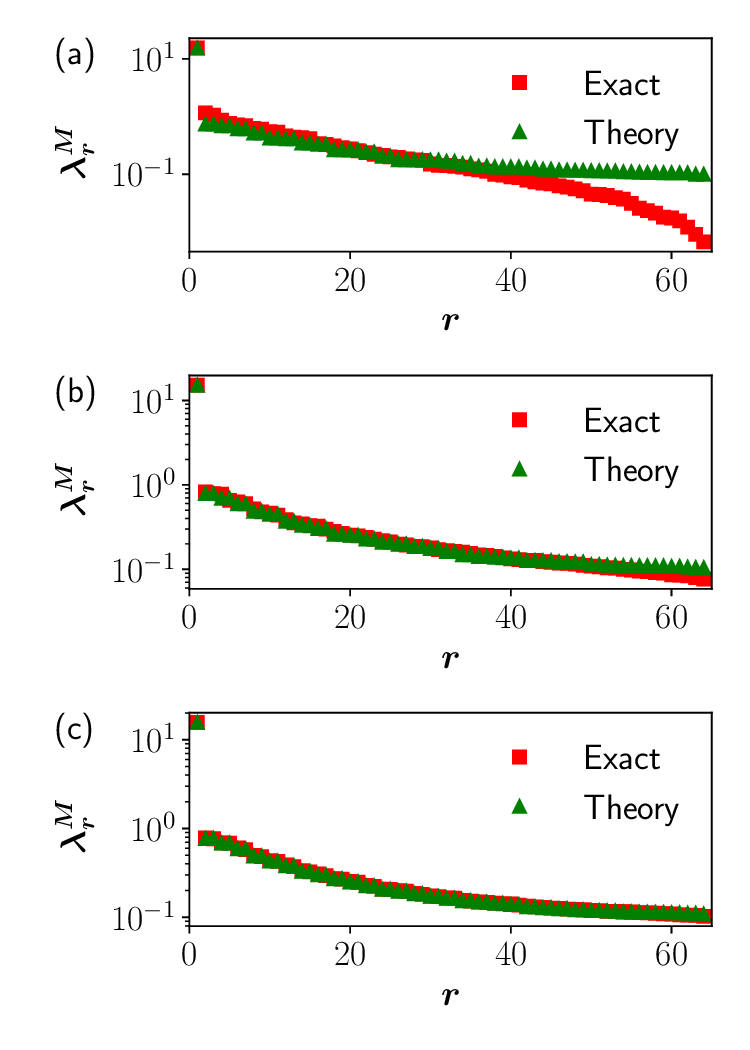}
\caption{\label{Fig:cp1} 
Comparison of $\lambda_r^M$ 
with Eq.~\eqref{Eq:theorya} for the one-dimensional CP with $L=64$ at $t=40$.
The exact eigenvalues (squares) and the
theoretical approximations (triangles) are drawn as a function
of ranking $r$ on a semilogarithmic scale for (a) $N=10^2$, (b) $10^3$, and (c) $10^4$.
}
\end{figure}
In Fig.~\ref{Fig:cp1}, we compare the exact eigenvalues of $M$ 
with the approximation~\eqref{Eq:theorya} for $N=10^2$, $10^3$, and $10^4$, top to bottom. Indeed,
the approximation becomes better and better as $N$ increases.
The result in each panel of Fig.~\ref{Fig:cp1} is from a single realization of 
$X$ for each $N$, so it may look noisy.

We now ponder on the physical meaning of Eq.~\eqref{Eq:theory2} for the CP in $d$ dimensions. Let 
$$
\rho := \frac1L \sum_{\bm{a} \in {\cal L}} x_{\bm{a}},
$$
which is the density.
As we have already discussed, $\rho$ is the order parameter.
Now, we will show that the largest eigenvalue of $S$ is related with 
the order parameter $\rho$.
Since $S_{\bm{ab}} \ge 0$ for all pairs of $\bm{a}$ and $\bm{b}$ (recall $x_{\bm{a}}\ge 0$) and
\begin{align*}
\sum_{\bm{b}\in{\cal L}} S_{\bm{a}\bm{b}} &= 
\sum_{\bm{c}\in {\cal L}} \left [ x_{\bm{a}+\bm{c}} \frac1L \sum_{\bm{b}\in{\cal L}} x_{\bm{b}+\bm{c}} \right ]_N \\
&=\sum_{\bm{c}\in {\cal L}} \left [ x_{\bm{a}+\bm{c}} \rho \right ]_N = L 
\left [ \rho^2 \right ]_N = \tau_{\bm{0}}^M,
\end{align*}
for all $\bm{a}$, 
the largest eigenvalue of $S$ is $\tau_{\bm{0}}^M$ 
by the Perron-Frobenius theorem.
Here, the subscript $\bm{0}$ signifies the zero reciprocal lattice vector.
Therefore, $\lambda_1^M$ should be well approximated by $ L \left [ \rho^2 \right ]_N$ for
sufficiently large $N$.

In the literature,
the normalized largest eigenvalue $\tilde \lambda_1$ defined as
\begin{align}
\label{Eq:tl1}
\tilde \lambda_1 := 
\frac{\lambda_1^M}{\sum_r \lambda_r^M}
=\frac{\lambda_1^M}{\Tr M}
\end{align}
is usually investigated.
Together with $x_{\bm{a}}^2= x_{\bm{a}}$, Eq.~\eqref{Eq:tr} gives $\Tr M = L [\rho]_N$,
and, therefore,
$$
\tilde \lambda_1 \approx \frac{[\rho^2]_N}{[\rho]_N}.
$$
Since the CP is self-averaging in that for all $t$
$\rho$ converges almost surely to $\langle \rho \rangle$ as
$L\rightarrow \infty$ (``thermodynamic'' limit, so to say),
$\tilde \lambda_1$ should be very close to $\langle \rho\rangle$ 
as long as finite-size effect is negligible.
That is, $\tilde \lambda_1$ in the thermodynamic limit is just the order parameter.
This conclusion is applicable to any self-averaging model 
with $x_{\bm{a}}^2 = x_{\bm{a}}$. Note that the models studied in
Ref.~\cite{muzzi2024} have this property and, therefore,
the above discussion explains why $\tilde \lambda_1$ 
shows the same critical behavior as the order parameter;
see Figs.~3 and 9 of Ref.~\cite{muzzi2024},
where the mathematical symbol $\pi_1^u$ is used for $\tilde \lambda_1$ of this paper.
Other successful observations in Ref.~\cite{muzzi2024} regarding phase transitions 
are also understandable in the same line as above, together with the scaling 
theory~\cite{H2000,O2004}. 

In arriving at Eq.~\eqref{Eq:theory2}, we implicitly assumed that 
all eigenvalues of $S$ are not degenerate, to use the non-degenerate perturbation
theory. In fact, $S$ has many degenerate eigenvalues, as we will show soon. 
Hence, it is worth while to discuss whether the degenerate perturbation theory could yield 
better approximation.

\begin{figure}
\includegraphics[width=\linewidth]{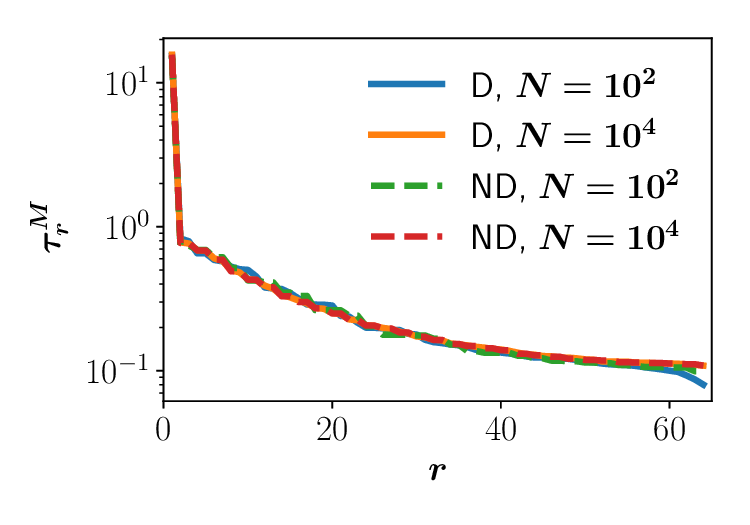}
\caption{\label{Fig:cp2} 
Comparison of the degenerate perturbation theory (solid lines)
with the non-degenerate one (dashed lines) for $N=10^2$ and $10^4$.
The degenerate perturbation theory gives hardly discernible result 
from Eq.~\eqref{Eq:theorya} at least for large $N$.
}
\end{figure}
For convenience, we treat the one-dimensional reciprocal lattice vector as 
a number; $\bm{p} = 2\pi k/L$ ($0\le k \le L-1$).
Since $\tilde x_{2\pi-\bm{p}}^* = \tilde x_{\bm{p}}$, 
we have the (at least) doubly degenerate eigenvalue 
$\tau_{\bm{p}}^{M}=\tau_{2\pi-\bm{p}}^{M}$ for $\bm{p} \neq 0, \pi$.
One may have observed this degeneracy already in Fig.~\ref{Fig:cp1}.
Using the degenerate perturbation theory, we get  ($\bm{p} \neq 0, \pi$)
\begin{align}
\tau_{\bm{p}\pm}^M 
= \tau_{\bm{p}}^M \pm \left | \langle 2\pi-\bm{p} | M | \bm{p} \rangle \right |
= \tau_{\bm{p}}^M \pm \left | \left [ {\tilde x_{\bm{p}}}^2 \right ]_N \right |.
\label{Eq:nond}
\end{align}
However, a numerical effort to calculate Eq.~\eqref{Eq:nond}
is not paid back especially for data with large $N$,
because $\delta M$ is very small anyway.
Indeed, as Fig.~\ref{Fig:cp2} shows,
almost no numerical effect of the degenerate perturbation theory 
is observed even for moderate size of samples.
Hence, we claim that the non-degenerate perturbation theory 
is already good even though the unperturbed matrix $S$ is highly degenerate.

Another interesting observation of Figs.~\ref{Fig:cp1} and \ref{Fig:cp2}
is that $\tau_r^M$ is less sensitive to $N$ than $\lambda_r^M$.
This observation has an important implication on practical limitations of the PCA 
in comparison to Eq.~\eqref{Eq:theory}, let alone
the computational complexity of calculating $\tau$, $O(NL\ln L)$ of the fast Fourier transform,
and calculating $\lambda$, $O(NL^2)$ of matrix multiplication and $O(L^3)$ of exact diagonalization of 
a non-sparse matrix.
The limitations show up conspicuously when we consider a large system.
Since, strictly speaking, a phase transition is defined in the thermodynamic limit, 
it is generally desirable to simulate a large system 
to avoid a finite-size effect. Due to the self-averaging property,
$\tau_{\bf{p}}^{M}$ even from a single (typical) realization of a very large system 
($L=10^{10}$, for instance) can be an accurate estimate of $\varphi_{\bf{p}}^{M}$.
In the PCA study with $N=1$, however, there is only one nonzero eigenvalue
$\lambda_1^M=\Tr M = \sum_{\bm{a}} x_{\bm{a}}^2$ and other $L-1$ eigenvalues are all zero 
(because $M$ for $N=1$ is a rank-one matrix).
For the CP, 
$\Tr M /L$ happens to be the empirical average of the order parameter $[\rho]_N$, so 
the PCA may be used to study the absorbing phase 
transition even with $N=1$ (although no one would study PCA with $N=1$).
But, this is a specialty of the CP. For the Ising model to be
studied in the next subsection, $\Tr M = L$ for all cases and a PCA study
with $N=1$, however large $L$ is, is completely pointless.
Since simulating a large system is demanding, 
to have many independent configurations for large $L$, which is a prerequisite of PCA, 
would be unfeasible. Hence, the PCA is practically limited to a smaller system
than Monte Carlo simulations can handle.

Let us move on to $\lambda_r^G$.  Since $[x_n]_N -[x_m]_N \aslim 0$ for all $1 \le n ,m\le L$ 
due to the translational invariance, 
we have $[\tilde x_{\bm{p}}]_N \aslim 0$ for nonzero ${\bm{p}}$. Therefore, we have
$\tau_{\bm{p}}^G - \tau_{\bm{p}}^M \aslim 0$ for nonzero ${\bm{p}}$. 
Only the Fourier mode with ${\bm{p}}=\bm{0}$ can remain different 
as one collects more and more samples.
Numerically, we indeed observed that $\lambda_r^G \approx \lambda_r^M$ for $r\ge 2$ and 
$\lambda_1^G \approx \tau_{\bm{0}}^G$ (details not shown here).
Since $\tau_{\bm{0}}^G = L ([\rho^2]_N - [\rho]_N^2 )$ is the fluctuation
of the order parameter that is supposed to diverge only at the critical point 
as $L\rightarrow\infty$, $G$ can also be used to study the 
absorbing phase transition in a different perspective from $M$.
\subsection{\label{Sec:ising}Two-dimensional Ising model}
As a second application,
we consider the two dimensional Ising model with Hamiltonian,
\begin{align}
\label{Eq:IH}
H(C) = - J \sum_{\langle {\bm{a}}, {\bm{c}}\rangle} \xi_{\bm{a}} \xi_{\bm{c}},
\end{align}
where $\langle \bm{a}, \bm{c}\rangle$ signifies the sum over nearest neighbor
pairs. For convenience, we assume $L_1 = L_2$ and $L=L_1^2$.
In this subsection, $\xi_{\bm{a}}$ takes either $+1$ or $-1$.

We first consider the ferromagnetic Ising model ($J>0$).
To sample $N$ configurations from the equilibrium 
distribution~\eqref{Eq:equil} at the critical point
$\beta_c J = \ln(1+\sqrt{2})/2$~\cite{Onsager1944}, 
we used the Metropolis algorithm with single-spin flip dynamics.
Since the dynamic exponent $z$ is about 2.168 
(see, e.g., Ref.~\cite{Park2015}), we sampled configurations
in every $L\times L_1^{2.2}$ attempts of spin flips, to guarantee
that configurations are independently sampled from the equilibrium distribution.
We sampled $10^4$ configurations for analysis. We constructed $X$, by setting $x_{\bm{a}}=
\xi_{\bm{a}}$. 

\begin{figure}
\includegraphics[width=\linewidth]{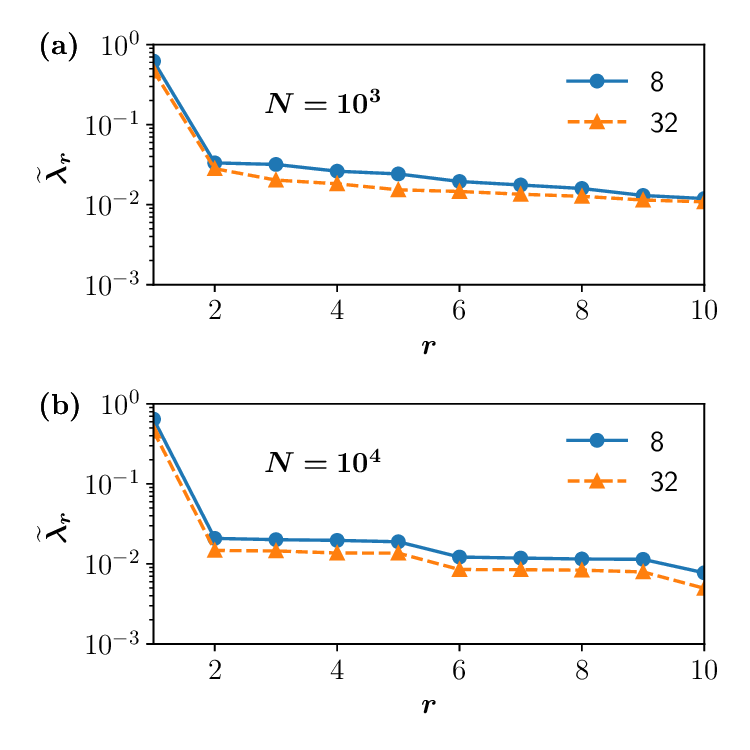}
\caption{\label{Fig:tl1} 
Semilogarithmic plots of $\tilde \lambda_r$ vs ranking $r$ of ten leading eigenvalues
for $L_1 = 8$ (circles) and $32$ (triangles). The number of configurations are 
(a) $N=10^3$ and (b) and $10^4$.
}
\end{figure}
At first, we present the numerical results of the normalized eigenvalues 
$\tilde \lambda_r := \lambda_r^M/\Tr M$ obtained by the matrix diagonalization. 
Notice that $\Tr M = L$ for any $N$ and for any temperature due to $x_{\bm{a}}^2=1$.
In Fig.~\ref{Fig:tl1}, we depict $\tilde \lambda_r$ against ranking $r$ for
$N=10^3$ and $10^4$, top to bottom. Note that Fig.~\ref{Fig:tl1}(a) looks consistent
with Fig.~1 of Ref.~\cite{Wang2016} that is obtained with $N=1400$ (mathematical
symbols $M$ and $\ell$ are actually used in Ref.~\cite{Wang2016} to denote 
$N$ and $r$ of this paper, respectively). 
Since the result in each panel of Fig.~\ref{Fig:tl1} is from a single
realization of $X$, the data may look noisier than Ref.~\cite{Wang2016} in which
an average over different realizations of $X$ is considered.

For $N = 10^3$, $\tilde \lambda_r$ shows steadily decreasing behavior with $r$,
whereas for $N=10^4$, $\tilde \lambda_r$ exhibits 
a somewhat regular step-like pattern with four similar values.
This almost four-fold degeneracy is understandable if Eq.~\eqref{Eq:theory2} is indeed valid.
Since the equilibrium distribution of the Ising model with $L_1=L_2$ is also invariant 
under the permutation of components of lattice vectors $(n_1,n_2) \mapsto (n_2,n_1)$, 
we have $\langle x_{(n_1,n_2)} x_{(n_3,n_4)}\rangle
=\langle x_{(n_2,n_1)} x_{(n_4,n_3)}\rangle$ and, accordingly,
\begin{align*}
 \left \langle \left | \tilde x_{\bm{p}} \right |^2 \right \rangle 
= \left \langle \left | \tilde x_{\bm{\bar p}} \right |^2 \right \rangle ,
\end{align*}
where $\bm{p}=(p_1,p_2)$ and $\bm{\bar p}=(p_2,p_1)$ are reciprocal lattice vectors
with $p_j = 2 \pi k_j /L_1$.
Hence, by the strong law of large numbers, we have
$$
\tau_{\bm{p}}^M
-\tau_{\bm{\bar p}}^M\aslim  0.
$$
Let $\bm{q} := 2\pi(1,1)$.
Since ${\tilde x}_{\bm{q} - \bm{p}} = \tilde x_{\bm{p}}^*$, 
we also have $\tau_{\bm{q}-\bm{p}}^M = \tau_{\bm{p}}^M$.
Therefore, $\tau^M_{\bm{p}} - \tau_{\bm{p}'}^M \aslim 0 $
for all $\bm{p},\bm{p}' \in \{(p_1,p_2), (p_2,p_1),
\bm{q} - (p_1,p_2),
\bm{q} - (p_2,p_1)\}$,
which is consistent with numerical observation in Fig.~\ref{Fig:tl1}(b).

\begin{figure}
\includegraphics[width=\linewidth]{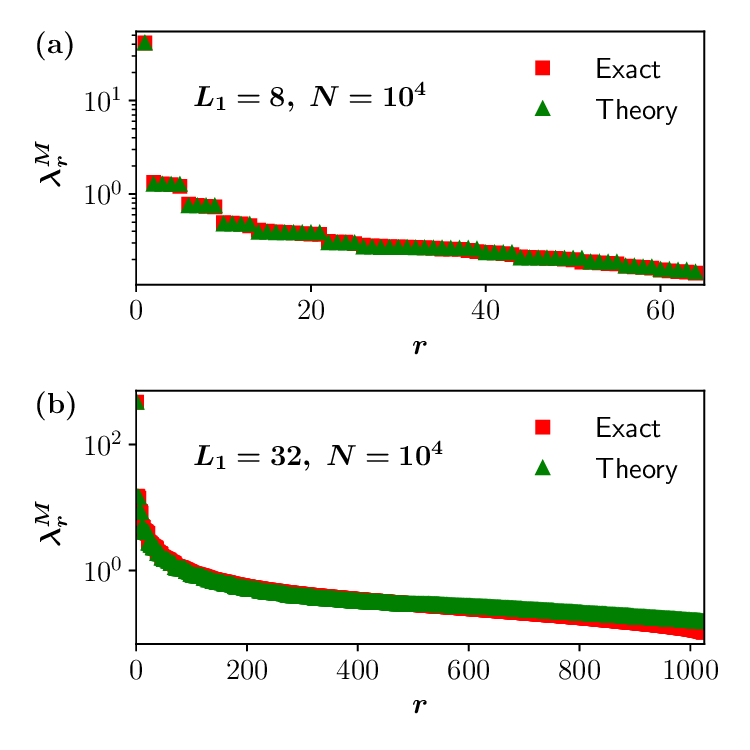}
\caption{\label{Fig:Ising} 
Semilogarithmic plots of eigenvalues of $M$ 
obtained by the matrix diagonalization (squares) and by the approximation
$\tau_r^M$ (triangles) against ranking $r$
for (a) $L_1 = 8$ and (b) $32$. The number of sampled configurations is $N=10^4$.
}
\end{figure}
Now we numerically support the validity of Eq.~\eqref{Eq:theory2}.
In Fig.~\ref{Fig:Ising}, we compare $\lambda_r^M$ with $\tau_r^M$
for $N=10^4$ with $L_1=8$ and $32$, top to bottom. Again, a good agreement
is observed. Since the system is homogeneous, it is not difficult to anticipate that 
$\tau_{\bm{0}}^M$ should be the largest among all $\tau$'s and therefore $\lambda_1^M \approx \tau_{\bm{0}}^M$.
Observing $\tau_{\bm{0}}^M = L [{\cal M}^2]_N$, 
where 
$$
{\cal M} := \frac1L\sum_n \xi_n
$$ 
is the magnetization,
we can conclude that the normalized largest eigenvalue $\tilde \lambda_1$   
is directly related to the order parameter~\cite{Wang2016,WZ2018}.

Following the discussion at the end of the previous subsection,
one can easily deduce that $\tau_{\bm{p}}^G$ should be very close to $\tau_{\bm{p}}^M$ for
$\bm{p}\neq (0,0)$, which was confirmed numerically (details not shown here). 
Since $\tau_{\bm{0}}^G = L\left ( [{\cal M}^2]_N - [{\cal M}]_N^2 \right )$ is 
related to the susceptibility,
$\lambda_1^G$, along with the finite-size scaling at criticality,
can also pinpoint the critical point.

Let us briefly discuss the stationary state of the Kawasaki dynamics~\cite{Kawasaki1966} 
of the two dimensional Ising model with zero magnetization. 
Since ${\cal M}=0$ by fiat, we always have $\varphi_{\bm{0}}^M=\tau_{\bm{0}}^M=0$.
Therefore, $\lambda_1^M$ should be related with a Fourier mode with nonzero reciprocal lattice vector.
Due to the permutation symmetry discussed above, there should be four significant principal components,
which was indeed observed in Fig. 4 of Ref.~\cite{Wang2016}.

The Ising model offers a good framework for exploring the role of $f$
in the definition $x_{\bm{a}} = f(\xi_{\bm{a}})$.
In the final part of this subsection, 
we present two examples to illustrate how a physically meaningful choice of $f$ can 
influence the PCA.

In the first example, 
we choose a lattice-vector dependent $f$ such as
$$
x^A_{(n_1,n_2)} = (-1)^{n_1+n_2} \xi_{(n_1,n_2)}
= e^{-i \pi (n_1+n_2)}\xi_{(n_1,n_2)},
$$
where the superscript $A$ is added for clarity.
We assume that both $L_1$ and $L_2$ are even.
Since $\xi$ is a spin of the ferromagnetic Ising model, 
it is obvious that $x^A$ is a spin of the antiferromagnetic Ising model with negative $J$
in the Hamiltonian~\eqref{Eq:IH}.
Therefore, $X$ made from $x^A$ can be regarded as a collection of independent equilibrium configurations
of the antiferromagnetic Ising model.
Now, we have
\begin{align*}
\tilde x^A_{\bm{p}} &= \frac1{\sqrt{L}} \sum_{n1,n2} e^{i(p_1n_1+p_2n_2)} x^A_{(n_1,n_2)}\\
&= \frac1{\sqrt{L}} \sum_{n1,n2} e^{i(p_1-\pi)n_1+i(p_2-\pi)n_2} \xi_{(n_1,n_2)}
=\tilde \xi_{\bm{p'}},
\end{align*}
where $\tilde \xi$ is the discrete Fourier transform of $\xi$ and
$$
\bm{p} = 2\pi\left ( \frac{k_1}{L_1},\frac{k_2}{L_2} \right ),\quad
\bm{p'}=\bm{p}-\pi(1,1).
$$
In $\bm{p'}$, the operation modulo $2\pi$ on each component is assumed.
Therefore, PCA applied to the antiferromagnetic Ising model gives ``identical'' result to
the ferromagnetic case, when both $L_1$ and $L_2$ are even numbers.
Since the leading principal component of the ferromagnetic case is associated with 
the Fourier mode with zero $\bm{p}$, $\lambda_1^M$ 
of the antiferromagnetic case is approximated as
$$
\tau_{(\pi,\pi)}^M = L \left [ {\cal M}_s^2 \right ]_N,
\, {\cal M}_s := \frac1L\sum_{n_1,n_2} (-1)^{n_1+n_2} x^{A}_{(n_1,n_2)},$$
where ${\cal M}_s$ is the staggered magnetization, the order parameter of the antiferromagnetic 
Ising model on a square lattice.
As this example shows, one may study other system by an appropriate choice of $f$ 
than the original one.

In the first example, we have seen that a nontrivial choice of $f$ may yield a good result
for a different system from the original one.
In the second example, we will observe that a ``wrong,'' albeit meaningful, choice 
of $f$ may hinder PCA from being useful. For the given configurations of the ferromagnetic Ising model, 
we can study a lattice-gas model by setting $x_n^{\mathrm{LG}} = (1+\xi_n)/2$, where the superscript $\mathrm{LG}$ 
is added to distinguish it from $x$ studied in the beginning of this subsection.
We now construct $X$ and $\overline{X}$ with $x_n^{\mathrm{LG}}$.
Since $(x_n^{\mathrm{LG}})^2 = x_n^{\mathrm{LG}}$, we have 
$\Tr M = L(1+[{\cal M}]_N)/2$ and
$$
\tau_{\bm{p}}^M = 
\frac{1}4
\begin{cases}
\left [ \left |\tilde \xi_{\bm{p}}\right |^2 \right ]_N,& 
\bm{p}\neq \bm{0},\\
L \left ( 1 + 2 \left [{\cal M}\right ]_N + \left [ {\cal M}^2 \right ]_N
\right ), & \bm{p}=\bm{0}.
\end{cases}
$$
Obviously, we have
$$
\tilde \lambda_1 \approx \frac12
\frac{1+2\left [{\cal M}\right ]_N + \left [ {\cal M}^2 \right ]_N }{
1+\left [{\cal M}\right ]_N},
$$
which is not a good quantity for a numerical study of the phase transition in
the Ising model. This $\tilde \lambda_1$ is especially detrimental to the Kawasaki dynamics with 
zero magnetization, because  
we always have $\tilde \lambda_1 = \frac12$ regardless of temperature
due to ${\cal M} = 0$.
Hence, the PCA with $X$ constructed in terms of the lattice-gas degrees of freedom would fail
to find the critical point. 
Given that the lattice-gas model is a well-established physical one,
this example is clearly not an artifact.
Conversely, if $\xi$ is the degree of freedom of the lattice-gas model and if one choose
$x = f(\xi)=2 \xi - 1$, then the PCA correctly classifies the phases of the Ising model.
As this example shows, an appropriate choice of $f$, that is, a physical intuition,
has to be added to study a phase transition with principal components of $X$.
Meanwhile, we do not claim that the PCA in terms of the lattice-gas model 
is completely useless.
To study $G$ with $x^{\mathrm{LG}}$ obviously gives the same answer as in the study of $G$ with $x$.
\subsection{$q$-state Potts model}
In this subsection, we discuss the $q$-state Potts model with Hamiltonian
$$
H(C) = - J \sum_{\langle \bm{a},\bm{b} \rangle} \delta_{\xi_{\bm{a}},\xi_{\bm{b}}} - H \sum_{\bm{a}\in{\cal L}}
\delta_{\xi_{\bm{a}},0},
$$
where $\xi_{\bm{a}}\in \{0, 1, \ldots, q-1\}$. We consider $H\rightarrow 0^+$ limit.
Assume that we have collected $N$ independent configurations from the equilibrium distribution of the 
Potts model.
If we choose $f$ to be an identity function such that $x_{\bm{a}}=f(\xi_{\bm{a}})=\xi_{\bm{a}}$,
the $N\times L$ matrix $X$ would take the form ($q=3$ is assumed)
$$
X = \begin{pmatrix} 
1&0&2&\cdots&1\\
0&2&1&\cdots&0\\
2&1&1&\cdots&1\\
\vdots&\vdots&\vdots&\vdots&\vdots\\
1&1&2&\cdots&2\\
\end{pmatrix}.
$$

Using the theory in Sec.~\ref{Sec:met},
we now argue that PCA with the above $X$ will fail to predict the phase transition.
First of all, the actual value of $\xi$ has no physical significance. Even if we choose
$\xi = 0, 1, 10^{23}$ for the three-state Potts model, the physical nature of
the phase transition is not affected. However, $X$ as well as its singular values changes 
drastically. 
Hence, the configuration itself is not a good quantity to study the phase transition of the Potts
model.
Second, the failure of the PCA with the above $X$ is already anticipated by the discussion
about the lattice-gas model in the previous subsection. Note that the lattice-gas 
model is identical to the two-state Potts model with $\xi = 0, 1$ (with an appropriate
rescaling of $J$).
Although the empirical covariance matrix $G$ for $q=2$ can still have potential to predict
the phase transition, $G$ for $q \ge 3$ is not useful to study the phase transition because
$\langle \xi_{\bm{a}} \rangle$ has no physical meaning and, therefore, the eigenvalues of 
$G$ have nothing to do with physical fluctuations.

For the PCA to be successfully applied to the $q$-state Potts model, we have to choose a proper $f$.
Since the order parameter is
$$
{\cal M}_q := \frac1L\sum_{\bm{a} \in {\cal L}} \left ( \delta_{\xi_{\bm{a}},0} -  \frac{1}{q} \right ),
$$
by the choice of $f(\xi) = \delta_{\xi,0} -  1/q$ we have 
$\tau_{\bm{0}}^M = N \left [ ({\cal M}_q)^2 \right ]_N$
as the largest eigenvalue of the empirical second moment matrix $M$. 
Now the PCA correctly predicts the phase transition. 
This example clearly shows that a physical intuition through the
choice of $f$ should be added for the PCA to work. 

\subsection{$n$-vector model}
To take the spontaneous symmetry breaking in the $n$-vector model into
account properly, 
we consider the $h\rightarrow 0^+$ limit 
for the Hamiltonian
\begin{align}
\label{Eq:nvecHam}
-\beta H(C) = K \sum_{\langle \bm{a,b}\rangle} \sum_{\mu=1}^n 
\xi_{\bm{a}}^\mu\xi_{\bm{b}}^\mu
+ h \sum_{\bm{a} \in {\cal L}}\xi_{\bm{a}}^{1},
\end{align}
where $\xi_{\bm{a}}^\mu$ is the $\mu$th component of the $n$-dimensional
continuous vector $\xi_{\bm{a}}$ with modulus one.
When we discuss the $n$-vector model in Sec.~\ref{Sec:an}, an element of $X$
is implicitly set to be a vector.
In an actual practice of the PCA, however, one would not use a vector
as an element of a matrix $X$. Rather, one would construct $X$ as 
an $N \times nL$ matrix with components
\begin{align}
\label{Eq:nvecX}
X_{\alpha, (\mu-1)L+m} = \xi_m^{\mu(\alpha)},
\end{align}
where $1 \le \alpha \le N$, $1 \le \mu \le n$, $1 \le m \le L$, and
$\xi_m^{\mu(\alpha)}$ is the value of $\xi_m^\mu$ in configuration
$C_\alpha$. 
Recall that $m$ is related with a lattice vector $\bm{a}$ by 
$m= v(\bm{a})$ for a given VI mapping $v$.
Indeed, in the PCA study of the $XY$ model in the literature~\cite{Hu2017}, 
usually studied is
an $N\times 2L$ matrix
\begin{align*}
\begin{pmatrix}
\cos\theta_1^{(1)}&\cos\theta_2^{(1)}&\cdots&\sin\theta_1^{(1)}&\sin\theta_2^{(1)}&\cdots\\
\cos\theta_1^{(2)}&\cos\theta_2^{(2)}&\cdots&\sin\theta_1^{(2)}&\sin\theta_2^{(2)}&\cdots\\
\vdots&
\vdots&
\vdots&
\vdots&
\vdots&
\vdots
\end{pmatrix},
\end{align*}
or with its columns permuted,
where $\cos\theta_m^{(\alpha)} = \xi_m^{1(\alpha)}$ and
$\sin\theta_m^{(\alpha)} = \xi_m^{2(\alpha)}$ in our terminology.
Since a permutation of columns of a matrix cannot alter the singular values, 
it is enough to study $X$ in Eq.~\eqref{Eq:nvecX}.
In this subsection, we will show that the approximate singular values of the $X$ 
are also obtained using Fourier modes of $\xi$.

To this end,
we introduce an $N\times L$ matrix $X_\mu$ and
an $L\times L$ matrix $M_\mu$ as
\begin{align*}
X_\mu &:=\begin{pmatrix}
\xi_1^{\mu(1)}&\xi_2^{\mu(1)}&\cdots\\
\xi_1^{\mu(2)}&\xi_2^{\mu(2)}&\cdots\\
\vdots&
\vdots&
\vdots
\end{pmatrix},\quad (X_\mu)_{\alpha m} = \xi_m^{\mu(\alpha)},\\
M_\mu &:= \frac1N X_\mu^\dag X_\mu,
\end{align*}
for all $1\le \mu \le n$. 
Notice that $X_\mu$ can be thought of as $X$ in Sec.~\ref{Sec:mss}
with $f$ to be a coordinate function, i.e., $x = f(\xi)=\xi^{\mu}$.
Hence, $M_\mu$ can be analyzed with the method in Sec.~\ref{Sec:an},
by introducing an $L\times L$ matrix $S_\mu$ 
with components
$$
(S_\mu)_{\bm{ab}} = \frac1L \sum_{\bm{c} \in {\cal L}} \left [ \xi_{\bm{a}+\bm{c}}^\mu 
\xi_{\bm{b}+\bm{c}}^\mu \right ]_N.
$$
The translational invariance along with the strong law of large numbers 
gives that $M_\mu - S_\mu$ becomes negligible for large $N$.

Using $X_\mu$'s, we can write
\begin{align*}
X &= \begin{pmatrix} X_1&X_2&\cdots&X_n \end{pmatrix},\\
M &= \frac1N X^\dag X =\frac1N
\begin{pmatrix} 
X_1^\dag X_1 & X_1^\dag X_2&\cdots&X_1^\dag X_n\\
X_2^\dag X_1 & X_2^\dag X_2&\cdots&X_2^\dag X_n\\
\vdots&\vdots&\ddots&\vdots\\
X_n^\dag X_1 & X_n^\dag X_2&\cdots&X_n^\dag X_n\end{pmatrix}.
\end{align*}
We also introduce a block diagonal matrix $S$ as
$$
S = \text{diag}\begin{pmatrix}S_1&S_2&\cdots&S_n\end{pmatrix}.
$$
Since the modulus of $\xi$ is one,
we have $\Tr M = \Tr S = L$. 
Note that the element of $(1/N)X_\mu^\dag X_\nu$ has the form
$\left [ \xi_m^{\mu} \xi_\ell^{\nu}\right ]_N$.
Since the Hamiltonian~\eqref{Eq:nvecHam} is invariant under 
$\xi_{\bm{a}}^{\nu} \mapsto - \xi_{\bm{a}}^{\nu}$ for all $\bm{a}$
and for each $\nu$ with $\nu \ge 2$,
we have
$\left \langle \xi_m^{\mu} \xi_\ell^{\nu}\right \rangle =-\left \langle \xi_m^{\mu} \xi_\ell^{\nu}\right \rangle = 0$
for any pair of $\mu$ and $\nu$ with $\mu \neq \nu$.
Accordingly, $M-S$ becomes a negligible perturbation for large $N$
and, therefore, we have all (approximate) $nL$ eigenvalues of $M$ 
as the empirical average of squared moduli of Fourier modes for $\xi_{\bm{a}}^\mu$.

Now we associate the leading principal component with the order parameter.
Let 
$$
{\cal M}_\mu = \frac{1}{L} \sum_{\bm{a}\in{\cal L}} \xi_{\bm{a}}^\mu,
\quad 
{\cal M} = \frac{1}{L} \sum_{\bm{a}\in{\cal L}} \xi_{\bm{a}}.
$$
Recall that $\xi_{\bm{a}}$ is a vector and so is ${\cal M}$.
If no conservation of magnetization is imposed on the $n$-vector model, 
the largest eigenvalue of $M_\mu$ for infinite $N$ should be $ L 
\langle ({\cal M}_\mu)^2 \rangle$.
In the disordered phase with $h=0$, the $O(n)$ symmetry gives
$\langle ({\cal M}_\mu)^2 \rangle = (1/n)\langle {\cal M}^2 \rangle$ for
every $\mu$,
where ${\cal M}^2 = \sum_{\nu=1}^n ({\cal M}_\nu)^2$.
In the ordered phase for large $L$, we expect
$\langle {\cal M}^2 \rangle \approx \langle ({\cal M}_1)^2 \rangle \gg
\langle ({\cal M}_\mu)^2 \rangle$ for $\mu \ge 2$.
Hence, regardless of whether we use $X$ in Eq.~\eqref{Eq:nvecX}
or use $X$ with vector components as in Sec.~\ref{Sec:an}, 
PCA applied to the $n$-vector model 
gives the order parameter as the largest eigenvalue of $M$.

\section{\label{Sec:sum}Summary}
Until now, we investigated the physical implications of applying principal component 
analysis to classical lattice systems with the translational invariance. 

By applying the perturbation theory, we showed that the eigenvalues of both the 
empirical second-moment matrix $M$ and the empirical covariance matrix $G$
can be approximated using the discrete Fourier transform of the system's configurations, 
as outlined in Eqs.~\eqref{Eq:theory} and \eqref{Eq:theory2}. 
This result establishes a clear connection between the 
principal components derived from the PCA and the Fourier modes of the system.
To assess the accuracy of this approximation, we invoke the strong law of large numbers, 
demonstrating that the approximation becomes particularly accurate 
when the size of the dataset is large (big data, so to speak). 

Our theory provides an interpretation of the PCA as a numerical method for calculating 
the ensemble average of the squared moduli of the Fourier transforms, specifically the 
$\varphi_{\bm{p}}$'s in Eq.~\eqref{Eq:exact}. 
Furthermore, we demonstrated that the success of the PCA in identifying the critical point 
of a phase transition depends on whether the 
$\varphi_{\bm{p}}$'s are closely related to the order parameter of the phase transition 
under study. 

To illustrate these ideas concretely, we numerically examined two systems:
the one-dimensional contact process and the two-dimensional Ising model. 
In both cases, we observed excellent agreement between the results obtained from the PCA 
and the predictions in Eq.~\eqref{Eq:theory2}. 
We further discussed the $q$-state Potts model and the $n$-vector model in the framework
of the theory developed in Sec.~\ref{Sec:met}.
These examples served to validate our 
theoretical framework and demonstrated its applicability to other systems. 

Finally, our theory suggests that for practical applications, 
calculating Eq.~\eqref{Eq:theory} 
directly is more efficient (and more informative for small size of data from a large system) 
than performing the full PCA procedure.

\begin{acknowledgments}
This work was supported by the National Research Foundation of Korea (NRF) grant funded by the  
Korea government (MSIT) (Grant No. RS-2023-00249949).
\end{acknowledgments}
\bibliography{park}
\end{document}